\documentclass{sig-alternate}

\usepackage{amsmath}
\usepackage{amssymb}
\usepackage{graphicx}

\def\QED{\mbox{\rule[0pt]{1.5ex}{1.5ex}}}
\def\endproof{\hspace*{\fill}~\QED\par\endtrivlist\unskip}
\newtheorem{theorem}{Theorem}

\newtheorem{lemma}{Lemma}

\newtheorem{proposition}{Proposition}

\begin{document}

\title{Cross-Layer Scheduling for Cooperative Multi-Hop Cognitive Radio Networks}

\numberofauthors{1}
\author{\alignauthor{Dongyue Xue, Eylem Ekici}\\
\affaddr{Department of Electrical and Computer Engineering, Ohio
State University,} \affaddr{Columbus, OH 43202, USA}
\email{xued@ece.osu.edu, ekici@ece.osu.edu}}

\maketitle

\begin{abstract}
The paper aims to design cross-layer optimal scheduling algorithms
for cooperative multi-hop Cognitive Radio Networks (CRNs), where
secondary users (SUs) assist primary user (PU)'s multi-hop
transmissions and in return gain authorization to access a share of
the spectrum. We build two models for two different types of PUs,
corresponding to elastic and inelastic service classes. For CRNs
with elastic service, the PU maximizes its throughput while
assigning a time-share of the channel to SUs proportional to SUs'
assistance. For the inelastic case, the PU is guaranteed a minimum
utility. The proposed algorithm for elastic PU model can achieve
arbitrarily close to the optimal PU throughput, while the proposed
algorithm for inelastic PU model can achieve arbitrarily close to
the optimal SU utility. Both algorithms provide deterministic
upper-bounds for PU queue backlogs. In addition, we show a tradeoff
between throughput/utility and PU's average end-to-end delay
upper-bounds for both algorithms. Furthermore, the algorithms work
in both backlogged as well as arbitrary arrival rate systems.
\end{abstract}



\vspace{1mm} 




\vspace{1mm}

\section{Introduction}
In traditional networks, spectrum bands or channels are allocated to
licensed users. However, such fixed spectrum assignment gives rise
to the spectrum under-utilization problem as was reported by Federal
Communication Commission (FCC) \cite{survey0}. Cognitive Radio
Networks (CRNs) \cite{survey1} have recently emerged as a technology
for unlicensed users, referred to as secondary users (SUs), to
opportunistically utilize the spectrum assigned to licensed users,
referred to as primary users (PUs). Researchers have been working on
optimizing data rate and throughput of CRNs in single-hop settings
\cite{CR1}-\cite{CR4}. However, these works are not readily
extendable to multi-hop CRNs, since multi-hop transmission requires
that the CR policies take into account scheduling and routing
issues.

Back-pressure scheduling algorithms with Lyapunov optimization tools
have been extensively investigated for generic wireless networks
\cite{Lyap0}\cite{Lyap1}. In addition to the seminal work
\cite{Lyap0}, distributed and low-complexity algorithms have been
proposed in the literature such as \cite{Lyap5}\cite{Lyap6}. This
technique has been applied to CRNs in \cite{CR6}-\cite{CR9}.
Specifically, in \cite{CR6}, an optimal cross-layer scheduling
algorithm have been proposed in a single-hop setting to maximize SU
throughput subject to PU collision constraints. This single-hop
setting is extended in \cite{CR7} where aggregated utility is
maximized subject to PU power constraints. In \cite{CR8}, a
cooperative CRN is considered to optimize PU and SU utility, where
SUs assist PU transmission in a two-hop relay scenario, which is not
readily extendable to generic multi-hop CRNs. A multi-hop CRN
scheduling algorithm is proposed in \cite{CR9}, without considering
cooperation between PUs and SUs. To the best of our knowledge, no
throughput/utility-optimal scheduling algorithms have been proposed
in the literature for cooperative multi-hop CRNs.

In this paper, we propose two optimal cross-layer scheduling
algorithms for a multi-hop cooperative CRN, where SUs relay data for
a PU pair to gain access to the licensed spectrum. These two
algorithms aim to solve the throughput/utility maximization problem
under the so-called \emph{inelastic} and \emph{elastic} PU models.
In the inelastic PU model, the PU pair is guaranteed a minimum
utility and the SU utility is maximized. In this model, we consider
an adaptive-routing scenario where the routes of the PU flow are not
determined \emph{a priori}, which is more general than a
fixed-routing scenario. In the elastic PU model, the PU throughput
is maximized using fixed routes while the SUs are guaranteed a
throughput proportional to the PU data that they relay.

Salient contributions of our work with respect to the literature can
be listed as follows: (1) Both inelastic and elastic algorithms can
achieve a throughput/utility arbitrarily close to the optimal
values. (2) The algorithms guarantee deterministically upper-bounded
finite buffer sizes for PU queues in the CRN. (3) We identify a
tradeoff between the throughput/utility and the average end-to-end
delay upper-bounds for PU data: the inelastic algorithm achieves a
PU delay upper-bound of order $O(\frac{N^2}{\epsilon})$, i.e.,
\emph{polynomial delay} \cite{delay3} is achieved, where $N$ denotes
the number of nodes involved in PU relay and $\epsilon$
characterizes the difference between the achieved utility and the
optimal utility; The elastic algorithm achieves \emph{order optimal
delay} \cite{delay1}\cite{delay2}, i.e., the delay is upper-bounded
by the first order of the number of hops in a route. (4) Both
algorithms are extended from a backlogged source model to a model
with arbitrary arrival rates at transport layer.

The rest of the paper is organized as follows: Section 2 introduces
the network and PU models for the cooperative multi-hop CRN. In
Section 3, we propose and analyze the inelastic algorithm. The
elastic algorithm and its performances are provided in Section 4. In
Section 5, we extend both algorithms to the model with arbitrary
arrival rates at transport layer. We conclude our work in Section 6.


\section{Network Model}
In this section, we first present the overall multi-hop cooperative
CRN model, followed by analysis of the two PU models.

\subsection{Overall Network Elements and Constraints}
In this paper, we consider a multi-hop cooperative CRN where SUs
relay PU data in return for the right to use the wireless spectrum.
The multi-hop cooperative CRN in question can be divided into two
subnetworks: a ``PU relay subnetwork'' and an ``SU subnetwork''. The
PU relay subnetwork is composed of one primary source node ($s_P$),
a corresponding primary destination node ($d_P$), and a set of SUs
$\mathcal{S}$ that relay the PU traffic between $s_P$ and $d_P$ over
possibly multiple hops, where $\mid {\mathcal S}\mid = N$. We assume
that the channel condition cannot support direct transmission
between the PU pair, and thus PU data will be solely relayed by
secondary nodes. Denote the node set of the PU relay subnetwork by
$\mathcal{N}=\{s_P,d_P\}\cup\mathcal{S}$. Denoting the set of links
in PU relay subnetwork as $\mathcal{L}$, we can represent the PU
relay subnetwork as ($\mathcal{N}$, $\mathcal{L}$). Note that our
model is readily extendable to the scenario of multiple PU pairs.

The SU subnetwork is composed of the set of SUs $\mathcal{S}$ that
participate in PU data relaying, and the set of their one-hop
secondary neighbors $\mathcal{S}'$ with which they communicate. For
notational simplicity, we assume that ${\mathcal S}\cap {\mathcal
S'}=\emptyset$ and that there is a distinct SU $l'\in\mathcal S'$
that corresponds to every SU $l\in\mathcal S$. Then, the SU
subnetwork is represented by
$(\mathcal{S}\cup\mathcal{S}',\mathcal{L}')$, where
$\mathcal{L}'=\{(l,l')\mbox{: }
l\in\mathcal{S},\,\,l'\in\mathcal{S}'\}$ is the set of links in the
SU subnetwork. Note that our analysis can readily be extended to
cases where ${\mathcal S}\cap { \mathcal S'}\neq\emptyset$.

Let $\mathcal{V}=\mathcal{L}\cup\mathcal{L}'$. Then the CRN topology
is represented by an interference graph
$G=(\mathcal{V},\mathcal{E})$ (or sometimes referred to as conflict
graph). There is an edge in $\mathcal{E}$ between two links in
$\mathcal{V}$ if the links interfere with each other when scheduled
simultaneously. Furthermore, let $\mu_{mn}$ be the scheduled link
rate for PU data over link $(m,n)\in\mathcal{L}$, and denote the
scheduled SU link rate as $s_l$ over link $(l,l')\in\mathcal{L'}$.
For analytical simplicity, we assume a scheduled link rate takes a
value from $\{0,1\}$. A link schedule represented by a vector
$((\mu_{mn})_{(m,n)\in\mathcal{L}},(s_l)_{l\in\mathcal{S}})\in\{0,1\}^{|\mathcal{L}|+N}$
is said to be \emph{feasible} iff any pair of two scheduled links
does not belong to the interference edge set
$\mathcal{E}$. 
Let $\mathcal{I}$ be the set of feasible link schedules. Then, a
feasible link scheduler chooses a feasible link schedule
$((\mu_{mn}(t))_{(m,n)\in\mathcal{L}},(s_l(t))_{l\in\mathcal{S}})\in\mathcal{I}$
for each time slot $t$. In addition, we assume that each node only
possesses one transceiver that can only send or receive data from
one neighbor node. Thus,
$\forall{n\in\mathcal{N}\backslash\{s_P\}}$, the following
inequality holds:
\begin{equation}\label{eq:13}
\sum_{j:(j,n)\in\mathcal{L}}\mu_{jn}(t)+\sum_{i:(n,i)\in\mathcal{L}}\mu_{ni}(t)+\textbf{1}_{\{n\in\mathcal{S}\}}s_n(t)\leq
1, \mbox{ } \forall t,
\end{equation}
where $\textbf{1}_{\{x\}}$ is the indicator function for event $x$.
Note that since $s_P$ is the sender of the PU pair, we must have
\begin{displaymath}
\sum_{n\in\mathcal{S}}\mu_{ns_P}(t)=0, \mbox{ }\forall t.
\end{displaymath}

In the following two subsections, we build two PU models
corresponding to different PU service classes, namely, inelastic PU
model and elastic PU model. In the inelastic PU model,
adaptive-routing scenarios are considered and we maximize SU utility
while PU is guaranteed a minimum utility. In the elastic model, we
assume fixed-routing scenarios and maximize the PU throughput while
SUs obtain a throughput proportional to the PU data that they relay.

\subsection{Queueing Structure and Constraints for Inelastic PU Model}
In the inelastic PU model, we denote by $U_n(t)$ the queue backlog
for PU packets at node $n\in\mathcal{N}$, where $U_{d_P}(t)=0$
$\forall t$. Let $Q_l(t)$ be the queue backlog for SU packets
corresponding to the SU pair associated with $l\in\mathcal{S}$. Now
we define the stability of a generic queue with queue backlog
$X(t)$: $X(t)$ is said to be \emph{stable} if
\begin{displaymath}
\limsup_{T\rightarrow\infty}\frac{1}{T}\sum_{t=0}^{T-1}\mathbb{E}\{X(t)\}<\infty.
\end{displaymath}
Therefore, the network is \emph{stable} if queues $U_n(t)$ and
$Q_l(t)$ are stable $\forall n\in\mathcal{N}$ and $\forall
l\in\mathcal{S}$.

For the time being, we assume that PU and SU traffics are backlogged
at the transport layer. Thus, a congestion controller is needed to
admit packets into the network layer. Let $\mu_{ps_P}(t)$ be the
admitted PU arrival rate in time slot $t$. Note that we can consider
$p$ in the subscript of $\mu_{ps_P}(t)$ as the virtual node
representing the PU transport layer and consider $(p,s_P)$ as the
virtual link from transport layer to source PU, so we construct a
new link set as $\mathcal{L}^c\triangleq\mathcal{L}\cup\{(p,s_P)\}$.
Let $A_l(t)$, $l\in\mathcal{S}$, be the admitted SU arrival rates to
the SU pair associated with secondary node $l$ in time slot $t$. We
assume $\mu_{ps_P}(t)\leq \mu_M$ and $A_l(t)\leq A_M$ $\forall
l\in\mathcal{S}$, where $\mu_M$ and $A_M$ are the upper-bounds for
admitted PU and SU arrival rates, respectively. For analytical
simplicity, we assume that admitted packets are added to the queues
at the end of time slot $t$.

From the above analysis, we can develop the queueing dynamics for
$U_n(t)$, $n\in\mathcal{N}\backslash \{d_P\}$, as follows:
\begin{equation}\label{eq:01}
U_n(t+1)\leq [U_n(t)-\sum_{i:(n,i)\in\mathcal{L}}\mu_{ni}(t)]^+
+\sum_{j:(j,n)\in\mathcal{L}^c}\mu_{jn}(t),
\end{equation}
where $[x]^+=\max\{x,0\}$ and
$\sum_{i:(n,i)\in\mathcal{L}}\mu_{ni}(t)$ stands for the scheduled
service rate. Note that (\ref{eq:01}) is an inequality since a
feasible scheduler can be designed independent of the queue backlog
information. Specifically, the inequality holds when the actual
arrival rate at node $n$ is less than the scheduled arrival rate
$\sum_{j:(j,n)\in\mathcal{L}^c}\mu_{jn}(t)$, i.e., some neighbor
node $j$ does not have packets for the scheduled transmission
$\mu_{jn}(t)=1$. Similarly, $Q_l(t)$, $l\in\mathcal{S}$, evolves as
follows:
\begin{equation}\label{eq:02}
Q_l(t+1)=[Q_l(t)-s_l(t)]^+ +A_l(t).
\end{equation}

We denote by $f(x)$ and $g_l(x)$ with $l\in\mathcal{S}$,
respectively, the PU and SU utility functions of the time-average
transmission rate. As convention, we assume that the utility
functions are positive-valued, concave, strictly increasing and
continuously differentiable, with $f(0)=0$ and $g_l(0)=0$ $\forall
l\in\mathcal{S}$. Examples of utility functions include $\theta'
log(1+x)$ and $\theta' x$, where $\theta'>0$ is a weight for the
utility functions. We assume the inelastic PU imposes a minimum
utility constraint $a_P$, i.e., the utility of the time-average PU
transmission rate must be greater than or equal to $a_P$.

According to \cite{Lyap0}\cite{Lyap1}, we define the capacity region
$\Lambda_I$ of the inelastic CRN as the closure of all feasible
arrival rate vectors consisting of an admitted PU arrival rate and
$N$ admitted SU arrival rates, where each feasible arrival rate
vector is stabilizable by some scheduler. Without loss of
generality, we assume that there exists an SU rate vector
$(r_l)_{l\in\mathcal{S}}$ such that
$(f^{-1}(a_P),(r_l)_{l\in\mathcal{S}})$ is strictly inside
$\Lambda_I$, where $f^{-1}(x)$ is the inverse function of the
utility function $f(x)$. To assist the analysis, we let
$(r_{l,\epsilon}^*)_{l\in\mathcal{S}}$ be a solution to the
following optimization problem:
\begin{displaymath}
\max_{(r_l)_{l\in\mathcal{S}}:(f^{-1}(a_P)+\epsilon,(r_l+\epsilon)_{l\in\mathcal{S}})\in\Lambda_I}\sum_{l\in\mathcal{S}}g_l(r_l),
\end{displaymath}
where $\epsilon>0$ can be chosen arbitrarily small. Then according
to \cite{Lyap2}, we have:
\begin{displaymath}
\lim_{\epsilon\rightarrow
0^+}\sum_{l\in\mathcal{S}}g_l(r_{l,\epsilon}^*)=\sum_{l\in\mathcal{S}}g_l(r_l^*),
\end{displaymath}
where $(r_l^*)_{l\in\mathcal{S}}$ is a solution to the following
optimization:
\begin{displaymath}
\max_{(r_l)_{l\in\mathcal{S}}:(f^{-1}(a_P),(r_l)_{l\in\mathcal{S}})\in\Lambda_I}\sum_{l\in\mathcal{S}}g_l(r_l).
\end{displaymath}

In Section 3, we will propose an algorithm that satisfies the PU
minimum utility constraint and can achieve SU utility arbitrarily
close to the optimal value $\sum_{l\in\mathcal{S}}g_l(r_l^*)$, with
a tradeoff between the SU utility and the average PU delay
upper-bound.


\subsection{Routing and Queueing Structure for Elastic PU Model}
For the elastic PU model, we consider a fixed multi-path routing
scenario, where the PU data transmission have $K$ loopless
pre-determined routes. We denote the path for the $k$-th route as
$P_k=(v_k^0,v_k^1,...,v_k^{H_k},v_k^{H_k+1})$, where $(H_k+1)$ is
the total number of hops in the PU relay subnetwork for route $k$,
where $v_k^m\in\mathcal{N}$, $\forall m\in\{0,1,...,H_k+1\}$,
$\forall k\in\{1,2,...,K\}$. Without loss of generality, we assume
that each node $l\in\mathcal{S}$ is in at least one of the $K$
routes, that is: $\forall l\in\mathcal{S}$, $\exists k,m$ s.t.
$v_k^m=l$. Note that we always have $v_k^0=s_P$ and
$v_k^{H_k+1}=d_P$, $\forall k\in\{1,2,...,K\}$. According to this
routing structure, we construct PU queues $U_m^k(t)$ along the nodes
in the $K$ routes, where $0\leq m \leq H_k+1$ and $1\leq k\leq K$.
Note that, since $v_k^{H_k+1}=d_P$, we have $U_{H_k+1}^k(t)=0$,
$\forall t$, $\forall k\in\{1,2,...,K\}$.

Similar to the inelastic model, we assume that PU and SU traffics
are backlogged at the transport layer. Let $\mu_{-1,0}^k(t)$ be the
admitted arrival rate from the PU transport layer to the source PU
that is scheduled to pass through the $k$-th route. Note that,
consistent with the elastic model, we assume that the sum of
admitted PU arrival rates over $K$ routes is upper-bounded by
$\mu_M$, i.e,
\begin{displaymath}
\sum_{k=1}^{K}\mu_{-1,0}^k(t)\leq \mu_M,\mbox{ }\forall t.
\end{displaymath}
In addition, we let $\lambda_k$, $k=\{1,2,...K\}$, be the
time-average of $\mu_{-1,0}^k(t)$. Let $\mu_{m,m+1}^k(t)$, $0\leq
m\leq H_k$, be the scheduled rate for the hop $(v_k^m,v_k^{m+1})$
along the $k$-th path. Thus, $U_m^k(t)$ evolves as follows:
\begin{equation}\label{eq:03}
U_m^k(t+1)\leq [U_m^k(t)-\mu_{m,m+1}^k(t)]^+ +\mu_{m-1,m}^k(t),
\mbox{ }0\leq m\leq H_k,
\end{equation}
where the inequality holds if $\mu_{m-1,m}^k(t)=1$ and
$U_{m-1}^k(t)=0$, $1\leq m\leq H_k$. Note that a link
$(m,n)\in\mathcal{L}$ can be a hop in multiple routes, and hence we
can only schedule the hop with rate $1$ on one such route in any
time slot.

Let $\rho_k$ be the reward for SUs when a packet is admitted to
route $k$, i.e., $\rho_k\mu_{-1,0}^k(t)$ packets will be admitted
simultaneously to the SU queues corresponding to the nodes $v_k^m$,
$1\leq m\leq H_k$. Here, we assume that $\rho_k\mu_{-1,0}^k(t)$
takes integer values. Note that our analysis is readily extendable
to fractional-valued $\rho_k\mu_{-1,0}^k(t)$ by constructing a
counter that only admits $\lfloor\rho_k\mu_{-1,0}^k(t)\rfloor$
packets, where $\lfloor x\rfloor$ is the floor function. Also note
that the analysis can be extended to delayed rewards, i.e., a reward
rate $\rho_k\mu_{-1,0}^k(t)$ is admitted to SU queues at $t+\tau'$,
where $\tau'$ is the delay in unit of time slots.

From the above analysis, the SU queueing dynamics for $Q_l(t)$ can
be expressed as follows:
\begin{eqnarray}\label{eq:04}
\begin{aligned}
&Q_l(t+1)&\\
=&[Q_l(t)-s_l(t)]^+
+\sum_{k=1}^K\sum_{m=1}^{H_k}\rho_k\mu_{-1,0}^k(t)\textbf{1}_{\{v_k^m=l\}}&\\
=&[Q_l(t)-s_l(t)]^+ +\sum_{k=1}^K
\rho_k\mu_{-1,0}^k(t)\textbf{1}_{\{\exists m \mbox{: } v_k^m=l\}},&
\end{aligned}
\end{eqnarray}
where the second equality holds since each route is a loop-free.

The network is \emph{stable} if queues $U_m^k(t)$ and $Q_l(t)$ are
stable $\forall m,k$ $\forall l$. Then, we define the capacity
region $\Lambda_E$ of the elastic CRN as the closure of all feasible
arrival rate vectors each stabilizable by some scheduler. Note that
a feasible arrival rate vector is in the form of
\begin{displaymath}
((\lambda_k)_{k\in\{1,2,...,K\}},(\sum_{k=1}^K\rho_k\lambda_k\textbf{1}_{\{\exists
m \mbox{: } v_k^m=l\}})_{l\in{\mathcal{S}}})
\end{displaymath}
where $(\lambda_k)_{k\in\{1,2,...,K\}}$ represents the PU arrival
rates per route and
$(\sum_{k=1}^K\rho_k\lambda_k\textbf{1}_{\{\exists m \mbox{: }
v_k^m=l\}})_{l\in{\mathcal{S}}}$ represents the SU arrival rates
according to the reward mechanism. To assist the analysis, we let
$(\lambda_{k,\epsilon}^*)_{k\in\{1,2,...,K\}}$ be a solution to the
following optimization problem:
\begin{displaymath}
\begin{aligned}
&\max_{(\lambda_k)_{k\in\{1,2,...,K\}}}\sum_{k=1}^K\lambda_k&\\
\mbox{s.t.
}&{(\lambda_k):((\lambda_k+\epsilon),(\sum_{k=1}^K\rho_k(\lambda_k+\epsilon)\textbf{1}_{\{\exists
m \mbox{: } v_k^m=l\}}))\in\Lambda_E}&
\end{aligned}
\end{displaymath}
where $\epsilon>0$ can be chosen arbitrarily small. Similarly,
according to \cite{Lyap2}, we have:
\begin{displaymath}
\lim_{\epsilon\rightarrow
0^+}\sum_{k=1}^K\lambda_{k,\epsilon}^*=\sum_{k=1}^K\lambda_k^*,
\end{displaymath}
where $(\lambda_k^*)_{k\in\{1,2,...K\}}$ is a solution to the
following optimization:
\begin{displaymath}
\begin{aligned}
&\max_{(\lambda_k)_{k\in\{1,2,...,K\}}}\sum_{k=1}^K\lambda_k&\\
\mbox{s.t.
}&{(\lambda_k):((\lambda_k),(\sum_{k=1}^K\rho_k\lambda_k\textbf{1}_{\{\exists
m \mbox{: } v_k^m=l\}}))\in\Lambda_E}&
\end{aligned}
\end{displaymath}

In Section 4, we will propose an algorithm that can achieve a PU
throughput arbitrarily close to the optimal value
$\sum_{k=1}^K\lambda_k^*$, with a tradeoff between the PU throughput
and average PU/SU delay upper-bound.

\section{Inelastic Algorithm for the CRN}
In this section, we first introduce two types of virtual queues and
their structures to assist the development of the inelastic
algorithm. The inelastic algorithm is then introduced in Subsection
3.2.

\subsection{Virtual Queues and Approaches}
We construct a virtual queue $U_p(t)$ at the PU transport layer with
the following queue dynamics:
\begin{equation}\label{eq:05}
U_p(t+1)=[U_p(t)-\mu_{ps_P}(t)]^+ +R(t),
\end{equation}
where $R(t)$ denotes the virtual arrival rate to $U_p(t)$ in time
slot $t$ which will be determined by the inelastic algorithm in the
next subsection. Furthermore, let $R(t)$ be upper-bounded by
$\mu_M$. When $U_p(t)$ is stable, we know from queueing theory that
the time-average admitted PU arrival rate $\mu$ satisfies:
\begin{equation}\label{eq:06}
\mu\triangleq\lim_{T\rightarrow\infty}\sum_{t=0}^{T-1}\mu_{ps_P}(t)\geq
r\triangleq\lim_{T\rightarrow\infty}\sum_{t=0}^{T-1}R(t).
\end{equation}
The virtual queue $U_p(t)$, along with $R(t)$, regulates the
admitted PU arrival rate in the inelastic algorithm, in an attempt
to guarantee an average end-to-end delay upper-bound, as will be
stated in detail in the next subsection.

We construct another virtual service queue $Z(t)$ at the PU source
node $s_P$ with the following queueing dynamics:
\begin{equation}\label{eq:07}
Z(t+1)=[Z(t)-R(t)]^+ +f^{-1}(a_P).
\end{equation}
When $Z(t)$ and $U_p(t)$ are stable, we have $f(\mu)\geq f(r)\geq
a_P$. Specifically, the minimum utility constraint imposed by PU is
satisfied when the two virtual queues are stable.

\subsection{Inelastic Algorithm}
We design a control parameter $q_M$ indicating the buffer size for
each PU queue in the CRN, with $q_M\geq\mu_M$. The optimal inelastic
algorithm consists of four parts, namely, SU congestion controller,
$R(t)$ controller, PU congestion controller and a link scheduler,
described as follows.

\textbf{1) SU Congestion Controller}:
\begin{equation}\label{eq:08}
\min_{0\leq A_l(t)\leq A_M}A_l(t)Q_l(t)-V_1g_l(A_l(t)), \mbox{
}\forall l\in\mathcal{S},
\end{equation}
where $V_1>0$ is a control parameter in the algorithm. Note that we
always have $A_l(t)Q_l(t)-V_1g_l(A_l(t))\leq0$ under the SU
congestion controller, since $A_l(t)=0$ is a valid candidate for the
admitted arrival rate.

\textbf{2) $R(t)$ Regulator}:
\begin{equation}\label{eq:09}
\min_{0\leq R(t)\leq\mu_M}R(t)(U_p(t)\frac{q_M-\mu_M}{q_M}-Z(t)).
\end{equation}
Specifically, when $U_p(t)\frac{q_M-\mu_M}{q_M}-Z(t)>0$, the virtual
rate $R(t)$ is set to zero; otherwise, $R(t)=\mu_M$.

\textbf{3) PU Congestion Controller}:
\begin{equation}\label{eq:10}
\max_{0\leq\mu_{ps_P}(t)\leq\mu_M}\mu_{ps_P}(t)(q_M-\mu_M-U_{s_p}(t)).
\end{equation}
Specifically, when $q_M-\mu_M-U_{s_p}(t)\leq 0$, the admitted PU
arrival rate $\mu_{ps_P}(t)$ is set to zero; Otherwise,
$\mu_{ps_P}(t)=\mu_M$.

\textbf{4) Link Rate Scheduler}:
\begin{equation}\label{eq:11}
\max\sum_{(m,n)\in\mathcal{L}}\mu_{mn}(t)\frac{U_p(t)}{q_M}(U_m(t)-U_n(t))+\sum_{l\in\mathcal{S}}Q_l(t)s_l(t),
\end{equation}
with the constraint
$\{(\mu_{mn}(t))_{(m,n)\in\mathcal{L}},(s_l(t))_{l\in\mathcal{S}}\}\in\mathcal{I}$.
Note that when $U_m(t)-U_n(t)\leq 0$, $(m,n)\in\mathcal{L}$, we set
$\mu_{mn}(t)=0$ according to (\ref{eq:11}).

The inelastic algorithm has the following property:
\begin{proposition}
\begin{equation}\label{eq:12}
U_n(t)\leq q_M, \mbox{ }\forall n\in\mathcal{N}.
\end{equation}
\end{proposition}
\proof{We can prove Proposition 1 by induction. Initially when
$t=0$, $U_n(0)=0$ $\forall n\in\mathcal{N}$. Now assume in time slot
$t$ we have $U_n(t)\leq q_M$ $\forall n\in\mathcal{N}$. In
the induction step, we consider two cases:\\
Case 1: $n=s_P$. If $U_{s_P}(t)\leq q_M-\mu_M$, then since the
admitted arrival rate to $U_{s_P}(t)$ is bounded by $\mu_M$, we have
$U_{s_P}(t+1)\leq U_{s_P}(t)+\mu_M\leq q_M$. Otherwise, we have
$U_{s_P}(t)>q_M-\mu_M$, and according to the PU congestion
controller (\ref{eq:10}) we have $\mu_{ps_P}(t)=0$, from which we
obtain $U_{s_P}(t+1)\leq U_{s_P}(t)\leq q_M$.\\
Case 2: $n\neq s_P$. If $U_n(t)\leq q_M-1$, then we have
$U_n(t+1)\leq U_n(t)+1\leq q_M$ according to (\ref{eq:13}) and the
queueing dynamics (\ref{eq:01}). Otherwise, we have $U_n(t)= q_M$
and $U_n(t)\geq U_m(t)$ $\forall m\in\mathcal{N}$, and according to
the link scheduler (\ref{eq:11}) we have $\mu_{jn}(t)=0$ $\forall j$
such that $(j,n)\in\mathcal{L}$, from which we obtain $U_n(t+1)\leq
U_n(t)= q_M$ by the queueing dynamics (\ref{eq:01}).

Therefore, $U_n(t+1)\leq q_M$ $\forall n\in\mathcal{N}$, i.e., the
induction step holds, and the proposition is proved.
\endproof}

Now we present the main results of the inelastic algorithm in
Theorem 1.
\begin{theorem}
Let $\epsilon>0$ be chosen arbitrarily small.
Given that
\begin{equation}\label{eq:14}
q_M> \frac{\mu_M^2+N+1}{\epsilon}+\mu_M,
\end{equation}
the inelastic algorithm ensures the following inequality on queue
backlogs:
\begin{equation}\label{eq:15}
\limsup_{T\rightarrow
\infty}\frac{1}{T}\sum_{t=0}^{T-1}\mathbb{E}\{\sum_{l\in\mathcal{S}}Q_l(t)+U_p(t)+Z(t)\}\leq
\frac{B_1+V_1g_M}{\delta_1},
\end{equation}
where $B_1\triangleq
\frac{1}{2}\mu_M^2+\frac{1}{2}(f^{-1}(\mu_M))^2+\frac{\mu_M^2(q_M-\mu_M)}{q_M}+\frac{1}{2}N+\frac{1}{2}NA_M^2+\frac{1}{2}
\mu_Mq_M(N+1)$, $\delta_1$ is chosen such that
$0<\delta_1<\frac{\epsilon(q_M-\mu_M)-\mu_M^2-N-1}{2q_M}$, and $g_M$
is defined as:
\begin{displaymath}
\begin{aligned}
g_M&\triangleq\limsup_{T\rightarrow\infty}\frac{1}{T}\sum_{t=0}^{T-1}\mathbb{E}\{\sum_{l\in\mathcal{S}}g_l(A_l(t))\}-\sum_{l\in\mathcal{S}}g_l(r_{l,\epsilon}^*)&\\
&\leq\sum_{l\in\mathcal{S}}(g_l(A_M)-g_l(r_{l,\epsilon}^*)). &
\end{aligned}
\end{displaymath}
Furthermore, the inelastic algorithm achieves:
\begin{equation}\label{eq:16}
\sum_{l\in\mathcal{S}}g_l(a_l)\geq
\sum_{l\in\mathcal{S}}g_l(r_{l,\epsilon}^*)-\frac{B_1}{V_1},
\end{equation}
where $a_l$ is defined as the time-average ensemble value of
$A_l(t)$:
\begin{displaymath}
a_l\triangleq
\liminf_{T\rightarrow\infty}\frac{1}{T}\sum_{t=0}^{T-1}\mathbb{E}\{A_l(t)\}\mbox{,
} l\in\mathcal{S}.
\end{displaymath}
\end{theorem}

\emph{Remark 1 (Network Stability)}: The inequalities (\ref{eq:12})
from Proposition 1 and (\ref{eq:15}) from Theorem 1 indicate that
the inelastic algorithm stabilizes the actual and virtual queues. As
an immediate result, the network is stable and the minimum utility
constraint is met. In addition, Proposition 1 ensures that the
actual PU queues are \emph{deterministically} bounded by the finite
buffer size $q_M$.

\emph{Remark 2 (Optimal Utility and Tradeoff with Delay): }The
inequality (\ref{eq:16}) gives the lower-bound of the SU utility the
inelastic algorithm can achieve. Since the constant $B_1$ is
independent of the control parameter $V_1$, the algorithm can
achieve a utility arbitrarily close to the optimal value\\
$\sum_{l\in\mathcal{S}}g_l(r_l^*)$ as $\epsilon$ is chosen
arbitrarily small, with a tradeoff in the PU buffer size $q_M$ which
is of order $O(\frac{N}{\epsilon})$ as shown in (\ref{eq:14}). By
Little's Theorem, the average end-to-end delay upper-bound is of
order $O(\frac{N^2}{\epsilon})$. Note that it is easy to verify that
when applied to a fixed-routing scenario, the inelastic algorithm
achieves an average end-to-end delay upper-bound of order
$O(\frac{H^2}{\epsilon})$, where $H$ denotes the number of hops in
the route.

\emph{Remark 3 (Complexity of Algorithm): }In the inelastic
algorithm, the SU congestion controller, the $R(t)$ regulator and
the PU congestion controller can operate locally at SU transport
layer and source PU. The link rate scheduler is essentially a
centralized maximal weight matching problem
\cite{Lyap0}\cite{Lyap7}. To reduce complexity of the link rate
scheduler, suboptimal algorithms can be developed to at least
achieve a fraction $\gamma$ of the optimal utility. These suboptimal
algorithms include the well-studied Greedy Maximal Matching (GMM)
\cite{Lyap6} algorithm with $\gamma=\frac{1}{2}$ and the maximum
weighted independent set (MWIS) problem such as GWMAX and GWMIN
proposed in \cite{graph1} with $\gamma=\frac{1}{\Delta}$, where
$\Delta$ is the maximum degree of the CRN topology.

\emph{Remark 4 (Distributed Implementation of the Link Scheduler)}:
Distributed implementation can be developed in much the same way as
in \cite{Lyap5} to achieve \emph{a fraction} of the optimal utility.
In order to achieve a utility arbitrarily close to the optimal value
with distributed implementation, we can employ random access
techniques \cite{csma1}\cite{csma2} in the link scheduler with
fugacities \cite{csma3} chosen as
exp$\{\frac{\alpha\bar{U}_p(t)[U_m(t)-U_n(t)]^+}{q_M}\}$ for link
$(m,n)\in\mathcal{L}$ and exp$\{\alpha Q_l(t)\}$ for an SU link
associated with $l\in\mathcal{S}$, where $\bar{U}_p(t)$ is a local
estimate of $U_p(t)$ and $\alpha$ is a positive weight. It can be
shown that the distributed algorithm can still achieve an average PU
end-to-end delay of order $O(\frac{N^2}{\epsilon})$ with the
time-scale separation assumption \cite{delay2}-\cite{csma1}. Due to
limited space, a detailed discussion is omitted.

We prove Theorem 1 in the next subsection.

\subsection{Proof of Theorem 1}
Before we proceed, we present Lemma 1 as follows to assist us in
proving Theorem 1.
\begin{lemma}
For any feasible rate vector
$(\theta,(r_l)_{l\in\mathcal{S}})\in\Lambda_I$, there exists a
stationary randomized algorithm \emph{SI} that stabilizes the
network with SU admitted arrival rate $A_l^{SI}(t)=r_l$, $\forall t$
$\forall l\in\mathcal{S}$, and PU admitted arrival rate
$\mu_{ps_P}^{SI}(t)=\theta$, $\forall t$, and schedule
$\{(\mu_{mn}^{SI}(t))_{(m,n)\in\mathcal{L}},(s_l^{SI}(t))_{l\in\mathcal{S}}\}$
independent of queue backlogs satisfying:
\begin{displaymath}
\mathbb{E}\{\sum_{i:(n,i)\in\mathcal{L}}\mu_{ni}^{SI}(t)-\sum_{j:(j,n)\in\mathcal{L}^c}\mu_{jn}^{SI}(t)\}=0,\mbox{
}\forall t,\mbox{ }\forall n\in\mathcal{N};
\end{displaymath}
\begin{displaymath}
\mathbb{E}\{s_l^{SI}(t)\}=r_l,\mbox{ }\forall t, \mbox{ }\forall
l\in\mathcal{S}.
\end{displaymath}
\end{lemma}

Note that it is not necessary for the randomized algorithm \emph{SI}
to provide finite buffer size or delay guarantees. Similar
formulations of stationary randomized algorithms and existence
proofs have been presented in \cite{CR7}\cite{Lyap0}-\cite{Lyap2},
so we omit the proof of Lemma 1 for brevity.

\emph{Remark 5: }According to the $SI$ algorithm in Lemma 1, we
assign the virtual input rate as
$R^{SI}(t)=\mu_{ps_P}^{SI}(t)=\theta$, $\forall t$. Hence, the time
average of $R^{SI}(t)$ satisfies $r^{SI}=\theta$. Note that
$(\theta,(r_l)_{l\in\mathcal{S}})$ can take values as
$(f^{-1}(a)+\frac{1}{2}\epsilon,(r_{l,\epsilon}^*)_{l\in\mathcal{S}})$
or
$(f^{-1}(a)+\epsilon,(r_{l,\epsilon}^*+\epsilon)_{l\in\mathcal{S}})$.

We define the queue vector $\textbf{Q}_I(t)$ as:
\begin{displaymath}
\textbf{Q}_I(t)=((U_n)_{n\in\mathcal{N}},(Q_l)_{l\in\mathcal{S}},U_p(t),Z(t))
\end{displaymath}
and define the Lyapunov function $L_I(\textbf{Q}_I(t))$ as follows:
\begin{displaymath}
\begin{aligned}
&L_I(\textbf{Q}_I(t))
\triangleq\frac{1}{2}\{\sum_{l\in\mathcal{S}}Q_l(t)^2+\frac{q_M-\mu_M}{q_M}U_p(t)^2&\\
&\qquad\qquad\qquad\qquad+Z(t)^2+\sum_{n\in\mathcal{N}}\frac{U_n(t)^2U_p(t)}{q_M}\},
\end{aligned}
\end{displaymath}
where the last term of the above Lyapunov function takes a similar
form as in \cite{Lyap8}\cite{Lyap9}. Then, the corresponding
Lyapunov drift is defined by
\begin{displaymath}
\Delta_I(t)\triangleq\mathbb{E}\{L_I(\textbf{Q}_I(t+1))-L_I(\textbf{Q}_I(t))|\textbf{Q}_I(t)\}.
\end{displaymath}

By squaring both sides of the queueing dynamics
(\ref{eq:01})(\ref{eq:02})(\ref{eq:05})(\ref{eq:07}) and through
algebra, we can obtain:
\begin{eqnarray}\label{eq:22}
\begin{aligned}
&\Delta_I(t)-V_1\sum_{l\in\mathcal{S}}\mathbb{E}\{g_l(A_l(t))|\textbf{Q}_I(t)\}&\\
\leq &B_1+\frac{\mu_M^2+N+1}{2q_M}U_p(t)&\\
&-\sum_{n\in\mathcal{N}}\mathbb{E}\{\frac{U_n(t)U_p(t)}{q_M}(\sum_{i:(n,i)\in\mathcal{L}}\mu_{ni}(t)&\\
&\qquad\qquad\qquad\qquad\qquad-\sum_{j:(j,n)\in\mathcal{L}^c}\mu_{jn}(t))|\textbf{Q}_I(t)\}&\\
&-\mathbb{E}\{Z(t)(R(t)-f^{-1}(a_P))|\textbf{Q}_I(t)\}&\\
&-\mathbb{E}\{\frac{(q_M-\mu_M)U_p(t)}{q_M}(\mu_{ps_P}(t)-R(t))|\textbf{Q}_I(t)\}&\\
&-\mathbb{E}\{\sum_{l\in\mathcal{S}}Q_l(t)(s_l(t)-A_l(t))|\textbf{Q}_I(t)\}&\\
&-V_1\sum_{l\in\mathcal{S}}\mathbb{E}\{g_l(A_l(t))|\textbf{Q}_I(t)\},&
\end{aligned}
\end{eqnarray}
where we also employ the following inequalities:
\begin{displaymath}
\begin{aligned}
&\sum_{n\in\mathcal{N}}\frac{U_n(t+1)^2U_p(t+1)}{q_M}&\\
\leq&(\frac{R(t)}{q_M}+\frac{U_p(t)}{q_M})\sum_{n\in\mathcal{N}}U_n(t+1)^2 &\\
\leq&\mu_Mq_M(N+1)+\frac{U_p(t)}{q_M}(\mu_M^2+N+1)+\frac{U_p(t)}{q_M}\sum_{n\in\mathcal{N}}U_n(t)^2&\\
-&2\frac{U_p(t)}{q_M}
\sum_{n\in\mathcal{N}}U_n(t)(\sum_{i:(n,i)\in\mathcal{L}}\mu_{ni}(t)-\sum_{j:(j,n)\in\mathcal{L}^c}\mu_{jn}(t))\}.&
\end{aligned}
\end{displaymath}

Through algebra, we find the equivalence of (\ref{eq:22}):
\begin{eqnarray}\label{eq:23}
\begin{aligned}
&\Delta_I(t)-V_1\sum_{l\in\mathcal{S}}\mathbb{E}\{g_l(A_l(t))|\textbf{Q}_I(t)\}&\\
\leq &B_1+\frac{\mu_M^2+N+1}{2q_M}U_p(t)+f^{-1}(a_P)Z(t)&\\
+&\sum_{l\in\mathcal{S}}\mathbb{E}\{A_l(t)Q_l(t)-V_1g_l(A_l(t))|\textbf{Q}_I(t)\}&\\
+&\mathbb{E}\{R(t)(\frac{(q_M-\mu_M)U_p(t)}{q_M}-Z(t))|\textbf{Q}_I(t)\}&\\
-&\mathbb{E}\{\mu_{ps_P}(t)\frac{U_p(t)}{q_M}(q_M-\mu_M-U_{ps_P}(t)) |\textbf{Q}_I(t)\}&\\
-&\mathbb{E}\{\sum_{l\in\mathcal{S}}Q_l(t)s_l(t)&\\
&\quad+\sum_{(m,n)\in\mathcal{L}}\mu_{mn}(t)\frac{U_p(t)}{q_M}(U_m(t)-U_n(t))
|\textbf{Q}_I(t)\}.
\end{aligned}
\end{eqnarray}
Note that the last four terms of the RHS of (\ref{eq:23}) are
minimized by the SU congestion controller (\ref{eq:08}), the $R(t)$
regulator (\ref{eq:09}), the PU congestion controller (\ref{eq:10}),
and the link scheduler (\ref{eq:11}), respectively, over a set of
feasible algorithms including the stationary randomized algorithm
$SI$ introduced in Lemma 1 and Remark 5. Then, we substitute into
the fourth and fifth terms of the RHS of (\ref{eq:23}) (i.e., the
third and fourth lines of (\ref{eq:23})) a stationary randomized
$SI$ with admitted arrival rate vector
$(f^{-1}(a)+\frac{1}{2}\epsilon,(r_{l,\epsilon}^*)_{l\in\mathcal{S}})$
, and we substitute into the last two terms the $SI$ with admitted
arrival rate vector
$(f^{-1}(a)+\epsilon,(r_{l,\epsilon}^*+\epsilon)_{l\in\mathcal{S}})$.
After the above substitutions, we obtain:
\begin{eqnarray}\label{eq:24}
\begin{aligned}
&\Delta_I(t)-V_1\sum_{l\in\mathcal{S}}\mathbb{E}\{g_l(A_l(t))|\textbf{Q}_I(t)\}&\\
\leq&B_1-\frac{\epsilon(q_M-\mu_M)-\mu_M^2-N-1}{2q_M}U_p(t)&\\
&-\epsilon\sum_{l\in\mathcal{S}}Q_l(t)-\frac{\epsilon}{2}Z(t)-V_1\sum_{l\in\mathcal{S}}g_l(r_{l,\epsilon}^*)&\\
\leq&B_1-\delta_1(\sum_{l\in\mathcal{S}}Q_l(t)+U_p(t)+Z(t))-V_1\sum_{l\in\mathcal{S}}g_l(r_{l,\epsilon}^*),
\end{aligned}
\end{eqnarray}
where the second inequality holds when the condition (\ref{eq:14})
in Theorem 1 is satisfied.

We take the expectation of both sides of (\ref{eq:24}) over
$\textbf{Q}_I(t)$ and take the time average on $t=0,1,...,T-1$,
which leads to
\begin{eqnarray}\label{eq:25}
\begin{aligned}
&\frac{\delta_1}{T}\sum_{t=0}^{T-1}\mathbb{E}\{\sum_{l\in\mathcal{S}}Q_l(t)+U_p(t)+Z(t)\}&\\
\leq&
B_1+\frac{V_1}{T}\sum_{t=0}^{T-1}\mathbb{E}\{\sum_{l\in\mathcal{S}}g_l(A_l(t))\}-V_1\sum_{l\in\mathcal{S}}g_l(r_{l,\epsilon}^*(t)).&
\end{aligned}
\end{eqnarray}

By taking limsup of $T$ on both sides of (\ref{eq:25}), we can prove
(\ref{eq:15}). We can prove (\ref{eq:16}) by taking the liminf of
$T$ on both sides of (\ref{eq:25}) and by employing the following
fact from the concavity of the SU utility functions:
\begin{displaymath}
\sum_{l\in\mathcal{S}}g_l(\mathbb{E}\{A_l(t)\})\geq
\sum_{l\in\mathcal{S}}\mathbb{E}\{g_l(A_l(t))\}.
\end{displaymath}

Therefore, Theorem 1 is proved.

\section{Elastic Algorithm for the CRN}
In this section, we design the optimal elastic algorithm composed of
two parts, namely, PU congestion controller and a hop/link
scheduler, described in Subsection 4.1. Note that according to the
fixed-routing structure in PU relay subnetwork introduced in
Subsection 2.1, when developing the scheduler, we focus on the
hop/link schedule\\
$((\mu_{m,m+1}^k(t))_{m,k},(s_l(t))_{l\in\mathcal{S}}))$ which is
composed of a PU hop schedule and an SU link schedule. Note that
each hop schedule $(\mu_{m,m+1}^k(t))_{m,k}$ corresponds to a PU
link schedule $(\mu_{mn}(t))_{(m,n)\in\mathcal{L}}$.

\subsection{Elastic Algorithm}
\textbf{1) PU Congestion Controller}:
\begin{eqnarray}\label{eq:17}
\begin{aligned}
&\min\sum_{k=1}^K\mu_{-1,0}^k(t)(\rho_k\sum_{l\in\mathcal{S}}Q_l(t)\textbf{1}_{\{\exists
m:\mbox{ } v_k^m=l\}}&\\
&\qquad\qquad\qquad\qquad+U_0^k(t)-V_2)&\\
&\mbox{s.t. } \sum_{k=1}^{K}\mu_{-1,0}^k(t)\leq \mu_M, &
\end{aligned}
\end{eqnarray}
where $V_2$ is a control parameter in the algorithm. For time slot
$t$, define
$k^*\triangleq\mbox{arg}\min_k(\rho_k\sum_{l\in\mathcal{S}}Q_l(t)\textbf{1}_{\{\exists
m:\mbox{ } v_k^m=l\}}+U_0^k(t))$. Specifically, from (\ref{eq:17}),
we set
\begin{displaymath}
\mu_{-1,0}^{k^*}(t)= \left\{\begin{aligned}
            & \mu_M, \mbox{ if }\rho_{k^*}\sum_{l\in\mathcal{S}}Q_l(t)\textbf{1}_{\{\exists
m:\mbox{ } v_{k^*}^m=l\}}+U_0^{k^*}(t)\leq V_2,&\\
            & 0,\quad\mbox{otherwise.}&
         \end{aligned} \right.
\end{displaymath}
For $k\neq k^*$, we set $\mu_{-1,0}^k(t)=0$.

\textbf{2) Hop/Link Scheduler}:
\begin{eqnarray}\label{eq:18}
\begin{aligned}
&\max\{\sum_{k=1}^K\sum_{m=0}^{H_k}\mu_{m,m+1}^k(t)(U_m^k(t)-U_{m+1}^k(t))&\\
&\qquad\quad+\sum_{l\in\mathcal{S}}Q_l(t)s_l(t)\},&\\
&\mbox{s.t.
}\{(\mu_{mn}(t))_{(m,n)\in\mathcal{L}},(s_l(t))_{l\in\mathcal{S}}\}\in\mathcal{I},
\end{aligned}
\end{eqnarray}
where the optimization is taken over all feasible\\
$((\mu_{m,m+1}^k(t))_{m,k}$,$(s_l(t))_{l\in\mathcal{S}}))$ and we
note that each hop schedule $(\mu_{m,m+1}^k(t))_{m,k}$ corresponds
to a PU link schedule\\
$(\mu_{mn}(t))_{(m,n)\in\mathcal{L}}$. From (\ref{eq:18}), when
$U_m^k(t)-U_{m+1}^k(t)\leq 0$, $m\in\{0,1,...,H_k\}$, we set
$\mu_{m,m+1}^k(t)=0$.

The elastic algorithm has the following property:
\begin{proposition}
$\forall m\in\{0,1,...,H_k\}$, $\forall k\in\{1,2,...,K\}$, the
following inequality holds:
\begin{equation}\label{eq:19}
U_m^k(t)\leq U_M\triangleq\mu_M+V_2.
\end{equation}
\end{proposition}
\proof{Similar to the proof of Proposition 1, we prove Proposition 2
by induction. Initially when $t=0$, $U_m^k(0)=0$ $\forall m$,
$\forall k$. Now assume in time slot $t$ we have $U_m^k(t)\leq U_M$,
$\forall m$, $\forall k$. In the induction step, we consider two
cases:\\
Case 1: $m=0$. Given any route $k$, if $U_0^k(t)\leq V_2$, then we
have $U_0^k(t+1)\leq U_0^k(t)+\mu_M\leq U_M$ according to queueing
dynamics (\ref{eq:03}), where we recall that
$\mu_{-1,0}^k(t)\leq\mu_M$ from the constraint in PU congestion
controller (\ref{eq:17}). Otherwise, we have $V_2<U_0^k(t)\leq U_M$,
and hence we have
\begin{displaymath}
\rho_k\sum_{l\in\mathcal{S}}Q_l(t)\textbf{1}_{\{\exists m:\mbox{ }
v_k^m=l\}}+U_0^k(t)>V_2,
\end{displaymath}
which induces $\mu_{-1,0}^k(t)=0$ from the PU congestion controller
(\ref{eq:17}), and it follows that $U_0^k(t+1)\leq U_0^k(t)\leq U_M$
by the queueing dynamics
(\ref{eq:03}).\\
Case 2: $m\in\{1,2,...,H_k\}$, for any given route $k$. If
$U_m^k(t)\leq U_M-1$, then we have $U_m^k(t+1)\leq U_m^k(t)+1\leq
U_M$ according to queueing dynamics (\ref{eq:03}). Otherwise, we
have $U_m^k(t)=U_M\geq U_{m-1}^k(t)$, and according to the hop/link
scheduler we have $\mu_{m-1,m}^k(t)=0$, from which we have
$U_m^k(t+1)\leq U_m^k(t)=U_M$ by the queueing dynamics
(\ref{eq:03}).

Therefore, $U_m^k(t+1)\leq U_M$ $\forall m\in\{0,1,...,H_k\}$,
$\forall k\in\{1,2,...,K\}$, i.e., the induction step holds, and the
proposition is proved.
\endproof}
As a complement to Proposition 2, recall that given route $k$, we
always have $U_{H_k+1}^k(t)=0$, $\forall t$.

Now we present the main results of the elastic algorithm in Theorem
2.
\begin{theorem}
Let $\epsilon>0$ be chosen arbitrarily small. The
elastic algorithm ensures the following inequality on queue
backlogs:
\begin{equation}\label{eq:20}
\limsup_{T\rightarrow
\infty}\frac{1}{T}\sum_{t=0}^{T-1}\mathbb{E}\{\sum_{l\in\mathcal{S}}Q_l(t)\}\leq
\frac{B_2+V_2B_R}{\delta_2},
\end{equation}
where $B_2\triangleq
\frac{1}{2}K(N+2)+\frac{1}{2}N+\frac{1}{2}N\mu_M^2\max_k\rho_k^2$,
$\delta_2\triangleq\epsilon\min_k\rho_k$, and $B_R$ is defined as:
\begin{displaymath}
\begin{aligned}
B_R&\triangleq\limsup_{T\rightarrow\infty}\frac{1}{T}\sum_{t=0}^{T-1}\mathbb{E}\{\sum_{k=1}^K\mu_{-1,0}^k(t)\}-\sum_{k=1}^K\lambda_{k,\epsilon}^*&\\
&\leq \mu_M-\sum_{k=1}^K\lambda_{k,\epsilon}^*. &
\end{aligned}
\end{displaymath}

Furthermore, the inelastic algorithm achieves:
\begin{equation}\label{eq:21}
\liminf_{T\rightarrow\infty}\frac{1}{T}\sum_{t=0}^{T-1}\sum_{k=1}^K\mathbb{E}\{\mu_{-1,0}^k(t)\}\geq
\sum_{k=1}^K\lambda_{k,\epsilon}^*-\frac{B_2}{V_2}.
\end{equation}
\end{theorem}

\emph{Remark 6 (Stability): } The inequalities (\ref{eq:19}) from
Proposition 2 and (\ref{eq:20}) from Theorem 2 indicate that PU and
SU queues are all stable, and hence is the CRN. In addition,
Proposition 2 ensures that PU queues maintained in each route are
\emph{deterministically} bounded by the finite buffer size $U_M$.

\emph{Remark 7 (Optimal Throughput and Tradeoff with Delay): } The
inequality (\ref{eq:21}) gives the lower-bound of the throughput the
elastic algorithm can achieve. Since the constant $B_2$ is
independent of the control parameter $V_2$, the algorithm can
achieve a PU throughput arbitrarily close to the optimal value
$\sum_{k=1}^K\lambda_k^*$ as $\epsilon$ can be chosen arbitrarily
small and $V_2$ can be chosen arbitrarily large, with the following
tradeoffs in PU and SU delay:
\begin{itemize}
\item
The PU buffer size $U_M$ is of order $O(V_2)$ as shown in
(\ref{eq:19}). By Little's Theorem, 
the PU's average end-to-end delay over any given route $k$ is of
order $O((H_k+1)V_2)$ which is bounded by the first order of $H_k$,
i.e., the algorithm has \emph{order-optimal delay} per route.
\item
From (\ref{eq:20}), the average SU buffer occupancy is of order
$O(\frac{N+V_2}{\epsilon})$. And so is the SU average delay by
Little's Theorem. The average SU delay upper-bound has an extra term
$\frac{1}{\epsilon}$ in order compared with the average PU delay.
\end{itemize}

\emph{Remark 8 (Employing Delayed Queue Information): } The PU
congestion controller (\ref{eq:17}) is performed at the source PU.
Thus, in order to account for the propagation delay of queue
information $(Q_l(t))_{l\in\mathcal{S}}$, we can replace $(Q_l(t))$
in (\ref{eq:17}) by $(Q_l(t-\tau))$, where $\tau$ is an integer
number that is larger than the maximum propagation delay from any
node to a source. It is not difficult to show that Theorem 2 still
holds with a different value of $B_2$, with similar proof techniques
as in \cite{Lyap9}\cite{Lyap10}.

We prove Theorem 2 in the next subsection.

\subsection{Proof of Theorem 2}
Before we proceed, we present Lemma 2 as follows to assist us in
proving Theorem 2.
\begin{lemma}
For any feasible rate vector
\begin{displaymath}
((\lambda_k)_{k\in\{1,2,...,K\}},(\sum_{k=1}^K\rho_k\lambda_k\textbf{1}_{\{\exists
m \mbox{: } v_k^m=l\}})_{l\in\mathcal{S}})\in\Lambda_E,
\end{displaymath}
there exists a stationary randomized algorithm \emph{SE} that
stabilizes the network with PU admitted arrival rates
$\mu_{-1,0}^{k,SE}(t)=\lambda_k$, $\forall t$ $\forall
k\in\{1,2,...,K\}$ and a hop/link schedule\\
$((\mu_{m,m+1}^{k,SE}(t))_{m,k}$,$(s_l^{SE}(t))_{l\in\mathcal{S}})$
independent of queue backlogs satisfying:
\begin{displaymath}
\mathbb{E}\{\mu_{m-1,m}^{k,SE}(t)-\mu_{m,m+1}^{k,SE}(t)\}=0, \mbox{
}\forall t,\mbox{ }\forall m,k;
\end{displaymath}
\begin{displaymath}
\mathbb{E}\{s_l^{SE}(t)\}=\sum_{k=1}^K\rho_k\lambda_k\textbf{1}_{\{\exists
m \mbox{: } v_k^m=l\}} ,\mbox{ }\forall t,\mbox{ }\forall
l\in\mathcal{S}.
\end{displaymath}
\end{lemma}

Similar to Lemma 1, it is not necessary for the randomized algorithm
\emph{SE} to provide finite buffer size or delay guarantees. For
brevity, we omit the proof of Lemma 2, and interested readers are
referred to \cite{CR7}\cite{Lyap0}-\cite{Lyap2} for details. Note
that $(\lambda_k)_{k\in\{1,2,...,K\}}$ can take values as
$(\lambda_{k,\epsilon}^*)_k$ and
$(\lambda_{k,\epsilon}^*+\epsilon)_k$.

We denote the queue vector
$\textbf{Q}_E(t)=((U_m^k)_{m,k},(Q_l)_{l\in\mathcal{S}})$ and define
the Lyapunov function $L_E(\textbf{Q}_E(t))$ as follows:
\begin{displaymath}
L_E(\textbf{Q}_E(t))
\triangleq\frac{1}{2}\{\sum_{k=1}^K\sum_{m=0}^{H_k}(U_m^k(t))^2
+\sum_{l\in\mathcal{S}}Q_l(t)^2\}
\end{displaymath}
Then, the corresponding Lyapunov drift is defined as
\begin{displaymath}
\Delta_E(t)\triangleq\mathbb{E}\{L_E(\textbf{Q}_E(t+1))-L_E(\textbf{Q}_E(t))|\textbf{Q}_E(t)\}.
\end{displaymath}

By squaring both sides of the queueing dynamics
(\ref{eq:03})(\ref{eq:04}), we can obtain:
\begin{displaymath}
\begin{aligned}
&\Delta_E-V_2\mathbb{E}\{\sum_{k=1}^K\mu_{-1,0}^k(t)|\textbf{Q}_E(t)\} &\\
\leq&\frac{1}{2}\sum_{k=1}^K\sum_{m=0}^{H_k}\mathbb{E}\{(\mu_{m,m+1}^k(t))^2+(\mu_{m-1,m}^k(t))^2&\\
&\quad-2U_m^k(t)(\mu_{m,m+1}^k(t)-\mu_{m-1,m}^k(t))|\textbf{Q}_E(t)\}&\\
+&\frac{1}{2}\sum_{l\in\mathcal{S}}\mathbb{E}\{s_l(t)^2+(\sum_{k=1}^K\rho_k\mu_{-1,0}^k(t)\textbf{1}_{\{\exists
m\mbox{:
}v_k^m=l\}})^2&\\
&\quad-2Q_l(t)(s_l(t)-\sum_{k=1}^K\rho_k\mu_{-1,0}^k(t)\textbf{1}_{\{\exists
m\mbox{: }v_k^m=l\}}) |\textbf{Q}_E(t)\}&\\
-&V_2\mathbb{E}\{\sum_{k=1}^K\mu_{-1,0}^k(t)|\textbf{Q}_E(t)\},&
\end{aligned}
\end{displaymath}
from which we obtain:
\begin{eqnarray}\label{eq:26}
\begin{aligned}
&\Delta_E-V_2\mathbb{E}\{\sum_{k=1}^K\mu_{-1,0}^k(t)|\textbf{Q}_E(t)\} &\\
\leq&B_2-V_2\mathbb{E}\{\sum_{k=1}^K\mu_{-1,0}^k(t)|\textbf{Q}_E(t)\}&\\
-&\sum_{k=1}^K\sum_{m=0}^{H_k}\mathbb{E}\{U_m^k(t)(\mu_{m,m+1}^k(t)-\mu_{m-1,m}^k(t))
|\textbf{Q}_E(t)\} &\\
-&\sum_{l\in\mathcal{L}}\mathbb{E}\{Q_l(t)(s_l(t)&\\
&\qquad\qquad\quad-\sum_{k=1}^K\rho_k\mu_{-1,0}^k(t)\textbf{1}_{\{\exists
m\mbox{: }v_k^m=l\}})|\textbf{Q}_E(t)\}. &
\end{aligned}
\end{eqnarray}
Through algebra, we find the equivalence of (\ref{eq:26}):
\begin{eqnarray}\label{eq:27}
\begin{aligned}
&\Delta_E-V_2\mathbb{E}\{\sum_{k=1}^K\mu_{-1,0}^k(t)|\textbf{Q}_E(t)\}&\\
\leq&B_2+\mathbb{E}\{\sum_{k=1}^K
\mu_{-1,0}^k(t)\times&\\
&\quad(\rho_k\sum_{l\in\mathcal{S}}Q_l(t)\textbf{1}_{\{\exists
m:\mbox{ } v_k^m=l\}}+U_0^k(t)-V_2)|\textbf{Q}_E(t)\}&\\
&-\mathbb{E}\{\sum_{k=1}^K\sum_{m=0}^{H_k}\mu_{m,m+1}^k(t)(U_m^k(t)-U_{m+1}^k(t))&\\
&\qquad\qquad\qquad\qquad\qquad+\sum_{l\in\mathcal{S}}Q_l(t)s_l(t)|\textbf{Q}_E(t)\},
\end{aligned}
\end{eqnarray}
where we employ the following fact that $\forall k\in\{1,2,...,K\}$:
\begin{displaymath}
\begin{aligned}
&\sum_{m=0}^{H_k}U_m^k(t)(\mu_{m,m+1}^k(t)-\mu_{m-1,m}^k(t))&\\
=&
\sum_{m=0}^{H_k}\mu_{m,m+1}^k(t)(U_m^k(t)-U_{m+1}^k(t))-\mu_{-1,0}^k(t)U_0^k(t).&
\end{aligned}
\end{displaymath}

Note that the second and third terms of the RHS of (\ref{eq:27}) are
minimized by the PU congestion controller (\ref{eq:17}) and the
hop/link scheduler (\ref{eq:18}), respectively, over a set of
feasible algorithms including the stationary randomized algorithm
$SE$ introduced in Lemma 2. Then, we substitute into the second term
of the RHS of (\ref{eq:27}) a stationary randomized $SE$ with
admitted PU arrival rate vector
$(\lambda_{k,\epsilon}^*)_{k\in\{1,2,...,K\}}$ and into the third
terms the $SE$ with admitted PU arrival rate vector
$(\lambda_{k,\epsilon}^*+\epsilon)_{k\in\{1,2,...,K\}}$. After the
above substitutions, we obtain:
\begin{eqnarray}\label{eq:28}
\begin{aligned}
&\Delta_E-V_2\mathbb{E}\{\sum_{k=1}^K\mu_{-1,0}^k(t)|\textbf{Q}_E(t)\}&\\
\leq&B_2-V_2\sum_{k=1}^K\lambda_{k,\epsilon}^*-\epsilon\sum_{l\in\mathcal{S}}Q_l(t)\sum_{k=1}^K\rho_k\textbf{1}_{\{\exists
m\mbox{: }v_k^m=l\}}&\\
\leq&B_2-V_2\sum_{k=1}^K\lambda_{k,\epsilon}^*-\delta_2\sum_{l\in\mathcal{S}}Q_l(t).
\end{aligned}
\end{eqnarray}

We take the expectation of both sides of (\ref{eq:28}) over
$\textbf{Q}_E(t)$ and take the time average on $t=0,1,...,T-1$,
which leads to
\begin{eqnarray}\label{eq:29}
\begin{aligned}
&\frac{\delta_2}{T}\sum_{t=0}^{T-1}\mathbb{E}\{\sum_{l\in\mathcal{S}}Q_l(t)\}&\\
\leq&
B_2+\frac{V_2}{T}\sum_{t=0}^{T-1}\mathbb{E}\{\sum_{k=1}^K\mu_{-1,0}^k(t)\}-V_2\sum_{k=1}^K\lambda_{k,\epsilon}^*.&
\end{aligned}
\end{eqnarray}

By taking limsup of $T$ on both sides of (\ref{eq:29}), we can prove
(\ref{eq:20}). By taking the liminf of $T$ on both sides of
(\ref{eq:29}), we can prove (\ref{eq:21}). Therefore, Theorem 2 is
proved.

\section{Arbitrary Arrival Rates at Transport Layer}
In the previous model description and algorithm development, we
assumed that PU and SU packets are backlogged at the transport
layer. In this section, we present optimal algorithms for the
inelastic and elastic PU models, respectively, for arbitrary arrival
rates at the transport layer.

At the transport layer, let $E_p(t)$ and $E_l(t)$,
$l\in\mathcal{S}$, be the PU and SU arrival rates at the beginning
of time slot $t$, respectively. We assume that $E_p(t)$ and $E_l(t)$
$\forall l\in\mathcal{S}$ are i.i.d. with respect to time. For
simplicity of analysis, we assume that the time average arrival rate
vector, formed by the PU and SU arrival rates, is in the exterior of
the capacity region, so that a congestion controller is needed. Let
$W_p(t)$ and $W_l(t)$, $l\in\mathcal{S}$, be the backlog of PU and
SU data at the transport layer. PU and SU buffer sizes at the
transport layer are denoted by $W_P$ and $W_S$, respectively. In the
following subsections, we present modified algorithms that can
handle arbitrary arrival rates at transport layer for inelastic and
elastic PU models.

\subsection{Inelastic Algorithm for Arbitrary Arrival Rates at Transport Layer}
In the inelastic PU model, recalling that $A_l(t)$ is the admitted
SU rate, we update the SU backlog $W_l(t)$ at the transport layer as
follows:
\begin{equation}\label{eq:30}
W_l(t+1)=\min\{[W_l(t)+E_l(t)-A_l(t)]^+,W_S\},\mbox{ }\forall
l\in\mathcal{S}.
\end{equation}
Note that $W_l(t)=0$ $\forall t$ $\forall l\in\mathcal{S}$ and
$W_S=0$ if there is no buffer at SU transport layer. Similarly, we
update the PU backlog $W_p(t)$ at the transport layer as follows:
\begin{equation}\label{eq:31}
W_p(t+1)=\min\{[W_p(t)+E_p(t)-\mu_{ps_P}(t)]^+,W_P\}.
\end{equation}
Note that $W_p(t)=0$ $\forall t$ and $W_P=0$ when there is no buffer
at PU transport layer.

Following the idea introduced in \cite{Lyap1}, we construct a
virtual queue $Y_p(t)$ at the PU transport layer with queueing
dynamics:
\begin{displaymath}
Y_p(t+1)=[Y_p(t)-R(t)]^++u_p(t),
\end{displaymath}
where $u_p(t)$ is an auxiliary variable associated with $Y_p(t)$.
Similarly, we construct a virtual queue $Y_l(t)$, $l\in\mathcal{S}$,
at the SU transport layer with queueing dynamics:
\begin{displaymath}
Y_l(t+1)=[Y_l(t)-A_l(t)]^++u_l(t),
\end{displaymath}
where $u_l(t)$ is an auxiliary variable associated with $Y_l(t)$,
with $u_l$ being its time-average. Note that when $Y_l(t)$ is
stable, the time-average of SU admitted rate $A_l(t)$ is greater
than or equal to $u_l$. Thus, when $Y_l(t)$ is stable, $\forall
l\in\mathcal{S}$, if we can ensure that
$\sum_{l\in\mathcal{S}}g_l(u_l)$ is arbitrarily close to the optimal
value $\sum_{l\in\mathcal{S}}g_l(r_l^*)$, so is the SU utility.

Instead of (\ref{eq:07}), the queue state $Z(t)$ is now updated as:
\begin{displaymath}
Z(t+1)=[Z(t)-u_p(t)]^++f^{-1}(a_P).
\end{displaymath}
Denote the time-average of $u_p(t)$ as $u_p$. Thus, when $Y_p(t)$,
$U_p(t)$ and $Z(t)$ are stable, we have $f(\mu)\geq f(r)\geq
f(u_p)\geq a_P$, where we recall that $\mu$ and $r$ are the
time-average values of $\mu_{ps_P}(t)$ and $R(t)$, respectively.
Specifically, when $Y_p(t)$, $U_p(t)$ and $Z(t)$ are stable, the PU
minimum utility constraint is met.

Now we provide the inelastic algorithm for arbitrary arrival rates
at the transport layer:

\textbf{1) SU Congestion Controller}:\\
$u_l(t)$ and $A_l(t)$, $l\in\mathcal{S}$, are updated as follows:
\begin{equation}\label{eq:34}
\min_{0\leq u_l(t)\leq A_M}u_l(t)Y_l(t)-Vg_l(u_l(t))
\end{equation}
\begin{eqnarray}\label{eq:35}
\begin{aligned}
&\qquad\qquad\qquad\min A_l(t)(Q_l(t)-Y_l(t))&\\
&\qquad\qquad\mbox{s.t. }0\leq A_l(t)\leq\min\{W_l(t)+E_l(t),A_M\}&
\end{aligned}
\end{eqnarray}
Note that (\ref{eq:34}) and (\ref{eq:35}) can be solved
independently and locally at every SU.

\textbf{2) $R(t)$ Regulator}:
\begin{equation}\label{eq:36}
\min_{0\leq u_p(t)\leq \mu_M}u_p(t)(Y_p(t)-Z(t))
\end{equation}
\begin{eqnarray}\label{eq:37}
\begin{aligned}
&\qquad\qquad\min R(t)(\frac{q_M-\mu_M}{q_M}U_p(t)-Y_p(t))&\\
&\qquad\qquad\mbox{s.t. }0\leq R(t)\leq\min\{W_p(t)+E_p(t),\mu_M\}&
\end{aligned}
\end{eqnarray}
Note that (\ref{eq:36}) and (\ref{eq:37}) can be solved
independently and locally at the source PU.

\textbf{3) PU Congestion Controller}:
\begin{displaymath}
\begin{aligned}
&\qquad\max\mu_{ps_P}(t)(q_M-\mu_M-U_{s_p}(t))&\\
&\mbox{s.t. } 0\leq\mu_{ps_P}(t)\leq\min\{W_p(t)+E_(p),\mu_M\}&
\end{aligned}
\end{displaymath}

\textbf{4) Link Rate Scheduler}: The link scheduler is the same as
(\ref{eq:11}) in Section 3.2.

It is not difficult to check that Proposition 1 still holds, and we
present the following theorem for the performance of the algorithm:
\begin{theorem}
Let $\epsilon>0$ be chosen arbitrarily small. Given that
$q_M> \frac{\mu_M^2+N+1}{\epsilon}+\mu_M$,
the inelastic algorithm ensures the following inequality on queue
backlogs:
\begin{displaymath}
\begin{aligned}
&\limsup_{T\rightarrow
\infty}\frac{1}{T}\sum_{t=0}^{T-1}\mathbb{E}\{\sum_{l\in\mathcal{S}}(Q_l(t)+Y_l(t))+U_p(t)+Y_p(t)+Z(t)\}&\\
&\leq \frac{B_3+V_1g_M}{\delta_3},&
\end{aligned}
\end{displaymath}
where $B_3\triangleq B_1+NA_M^2+\mu_M^2$ and $\delta_3$ is a
constant. Furthermore, the algorithm achieves:
\begin{displaymath}
\sum_{l\in\mathcal{S}}g_l(a_l)\geq
\sum_{l\in\mathcal{S}}g_l(r_{l,\epsilon}^*)-\frac{B_3}{V_1},
\end{displaymath}
where we recall that $(a_l)_{l\in\mathcal{S}}$ is defined in Theorem
1.
\end{theorem}

The proof follows similar steps as the proof of Theorem 1. Due to
space limitations, we omit the proof for Theorem 3 and the choice of
$\delta_3$. Similar statements as presented in Remarks 1-4 also hold
for the inelastic algorithm introduced in this subsection.

\subsection{Elastic Algorithm for Arbitrary Arrival Rates at Transport Layer}
In this subsection, the elastic algorithm for arbitrary arrival
rates at transport layer and its performance are discussed. Similar
to inelastic algorithm, we denote by $A_l(t)$ the admitted SU rate,
which is upper-bounded by $A_M$. We update the SU backlogs $W_l(t)$
at the transport layer as (\ref{eq:30}). Similarly, we can update
the PU backlog $W_p(t)$ at the transport layer as:
\begin{displaymath}
W_p(t+1)=\min\{[W_p(t)+E_p(t)-\sum_{k=1}^K\mu_{-1,0}^k(t)]^+,W_P\}.
\end{displaymath}

Similar to the previous subsection, we construct virtual queues
$Y_p(t)$ and $Y_l(t)$, $\forall l\in\mathcal{S}$, with an auxiliary
variable $u_p(t)$ associated with $Y_p(t)$. The virtual queues
evolve as follows:
\begin{displaymath}
\begin{aligned}
&Y_p(t+1)=[Y_p(t)-\sum_{k=1}^K\mu_{-1,0}^k(t)]^++u_p(t);&\\
&Y_l(t+1)=[Y_l(t)-A_l(t)]^+&\\
&\qquad\qquad\quad+\sum_{k=1}^K\rho_k\mu_{-1,0}^k(t)\textbf{1}_{\{\exists
m\mbox{: }v_k^m=l\}}, \mbox{ }\forall{l\in\mathcal{S}}.&
\end{aligned}
\end{displaymath}
Note that when $Y_p(t)$ is stable, if we can ensure that $u_p$ is
arbitrarily close to the optimal value $\sum_{k=1}^K\lambda_k^*$,
then so is the PU throughput, where we recall that $u_p$ is the time
average of $u_p(t)$. In addition, when $Y_l(t)$ is stable, $\forall
l\in\mathcal{S}$, the SUs' throughput is proportional to the PU data
that they relay.

Now we provide the elastic algorithm for arbitrary arrival rates at
the transport layer:

\textbf{1) SU Congestion Controller}:\\
\begin{displaymath}
\min_{0\leq A_l(t)\leq
\min\{W_l(t)+E_l(t),A_M\}}A_l(t)(Q_l(t)-Y_l(t))\mbox{, }\forall
l\in\mathcal{S}.
\end{displaymath}
Note that the SU congestion controller can be solved locally at each
SU.

\textbf{2) PU Congestion Controller}:
\begin{equation}\label{eq:32}
\min_{0\leq u_p(t)\leq\mu_M}u_p(t)(Y_p(t)-V_2)\qquad\qquad
\end{equation}
\begin{eqnarray}\label{eq:33}
\begin{aligned}
&\quad\min\sum_{k=1}^K\mu_{-1,0}^k(t)(\rho_k\sum_{l\in\mathcal{S}}Y_l(t)\textbf{1}_{\{\exists
m:\mbox{ } v_k^m=l\}}&\\
&\qquad\qquad\qquad\qquad\qquad+U_0^k(t)-Y_p(t))&\\
&\mbox{s.t. } 0\leq\sum_{k=1}^{K}\mu_{-1,0}^k(t)\leq
\min\{W_p(t)+E_l(t),\mu_M\}. &
\end{aligned}
\end{eqnarray}
Note that (\ref{eq:32}) and (\ref{eq:33}) can be solved
independently.

\textbf{3) Link/Hop Rate Scheduler}: The link/hop scheduler is the
same as (\ref{eq:18}) in Section 4.1.

We present Proposition 3 and Theorem 4 to characterize the
performance of the proposed algorithm:
\begin{proposition}
$\forall m\in\{0,1,...,H_k\}$, $\forall k\in\{1,2,...,K\}$, the
following inequality holds:
\begin{displaymath}
U_m^k(t)\leq2\mu_M+V_2.
\end{displaymath}
\end{proposition}

\begin{theorem}
Let $\epsilon>0$ be chosen arbitrarily small. The
algorithm ensures the following inequality on queue backlogs:
\begin{displaymath}
\limsup_{T\rightarrow
\infty}\frac{1}{T}\sum_{t=0}^{T-1}\mathbb{E}\{\sum_{l\in\mathcal{S}}(Q_l(t)+Y_l(t))+Y_p(t)\}\leq
\frac{B_4+V_2B_R}{\delta_4},
\end{displaymath}
where $B_4\triangleq B_2+\mu_M^2+NA_M^2$ and $\delta_4>0$ is a
constant.

Furthermore, the inelastic algorithm achieves:
\begin{displaymath}
\liminf_{T\rightarrow\infty}\frac{1}{T}\sum_{t=0}^{T-1}\sum_{k=1}^K\mathbb{E}\{\mu_{-1,0}^k(t)\}\geq
\sum_{k=1}^K\lambda_{k,\epsilon}^*-\frac{B_4}{V_2}.
\end{displaymath}
\end{theorem}

The proofs of Proposition 3 and Theorem 4 follow similar steps as
the proofs of Proposition 2 and Theorem 2, respectively. Detailed
proofs and the choice of $\delta_4$ are omitted due to limited
space. Remarks 6-8 hold for the elastic algorithm introduced in this
subsection with minor modifications.

\section{Conclusions and Future Works}
In this paper, two cross-layer scheduling algorithms for multi-hop
cooperative cognitive radio networks are introduced. The algorithms
can achieve arbitrarily close to the optimal throughput/utility,
with a tradeoff in the deterministically upper-bounded PU buffer
sizes and hence the average end-to-end delay upper-bounds. Our work
aims at a better understanding of the fundamental properties and
performance limits of QoS-constrained multi-hop CRNs. In our future
work, we will investigate distributed implementations and power
management in CRNs.

\bibliographystyle{abbrv}

\end{document}